\documentclass{article}
\usepackage{graphicx}
\usepackage[english]{babel}
\usepackage{amsmath,amssymb}
\usepackage{siunitx}
\usepackage{color}
\usepackage[parfill]{parskip}

\DeclareMathOperator*{\argmin}{arg\,min}
\renewcommand{\vec}{\mathbf}

\begin{document}

\title{Effects of environment knowledge in evacuation scenarios involving fire and smoke - a multiscale modelling and simulation approach}

\author{Omar Richardson\thanks{Karlstad University, Sweden} \and Andrei Jalba\thanks{Eindhoven University of Technology, The Netherlands} \and Adrian Muntean\footnotemark[1]}

\date{June 8th, 2018}

\maketitle

\begin{abstract}
We study the evacuation dynamics of a crowd evacuating from a complex geometry in the presence of a fire as well as of a slowly spreading smoke curtain. The crowd is composed of two kinds of individuals: those who know the layout of the building, and those who do not and rely exclusively on potentially informed neighbors to identify a path towards the exit.

We aim to capture the effect the knowledge of the environment has on the interaction between evacuees and their residence time in the presence of fire and evolving smoke. Our approach is genuinely multiscale -- we employ a two-scale model that is able to  distinguish between compressible and incompressible pedestrian flow regimes and allows for \emph{micro} and \emph{macro} pedestrian dynamics. Simulations illustrate the expected qualitative behavior of the model. We finish with observations on how mixing evacuees with different levels of knowledge impacts important evacuation aspects.
\end{abstract}
\clearpage
\section{Introduction}
\label{sec:intro}
Fire safety of buildings has always been an important aspect of societal research. 
An important development in recent years is the shift from prescriptive building codes to performance-based building (\cite{vanhees13}), meaning that instead of requiring aspects of buildings to satisfy certain regulations, new building structures increasingly have to be able to attain certain goals, without exactly specifying how these can be accomplished. 
This increases the need of understanding the different aspects of fire emergencies.
Models and simulations of fire spread and evacuation scenarios play an important role in this trend, since they are able to complement experimental research.
This paper aims to contributes in models of the latter category: how and under what circumstances occupants evacuate from a building.

Building evacuations are too complex to be investigated exhaustively in short time. Differences in psychological state, rational, panicked or altruistic behaviour have significant impact on the progress of the evacuation.
Moreover, not all occupants share the same degree of familiarity with the building they are in, which greatly affects their behaviour.
In this paper, we focus on the interaction between two groups of occupants; one group is very familiar with the building and its layout and one group  misses this familiarity and must thus rely on the former group for evacuating safely.
One example of its relevance is described in \cite{horiuchi86} - an investigation of an evacuation from an office building, where, in hindsight, evacuees where asked about their knowledge of the building and their actions regarding the evacuation. Among the conclusions drawn was that there were significant behavioural differences between habitual occupants and one-time guests.
Not including this feature in crowd simulations can, in some cases, lead to large deviations from realistic evacuation behaviour. This is precisely the place where we contribute. 

We concern ourselves with a scenario resembling the one described in \cite{horiuchi86}: occupants with different levels of knowledge evacuate from a large indoor building.
By placing the occupants in a simulation-based framework including complex geometries, we computationally explore the effects the environment knowledge has on the evacuation speed and pedestrian flow congestion.
We stress the importance of knowledge by modifying the evacuation conditions to include fire and smoke production, limiting agents visual perception and freedom of movement.

This paper is the start of a larger initiative designed to develop interactive simulations as a decision support tool for crowd management, in which interactive evacuations can be steered and optimized based on real-time feedback. 
For this reason, any simulations proposed should be executable fast in different configurations and complex domains. To achieve this goal, we design a multiscale model that is averaged and fast on a macroscale as well as accurate and slow on a  microscale, but, for large crowds, it performs much faster than the usual agent-based models and is significantly more accurate than the purely macroscopic transport models.
By screening the crowd both in agent-based representation and continuum representation, we avoid costly agent-to-agent based interactions by deferring the counting of interactions to a macro (observable) scale.

\section{Related contributions}

In line with Pinter-Wollman et al. in \cite{pinter17}, we show in this paper that knowledge of the building layout drastically affects the collective behaviors of occupants walking through smoke towards the exits.
Several other related contributions exist. For instance, \cite{bae15} reports on the effect of smoke using a social force model \`a la Helbing.
Cirillo et al. (cf. \cite{cirillo13}, e.g.) look into the effects of limited visibility on the group dynamics during an evacuation from a (dark) room without obstacles. Their work has been expanded in \cite{cirillo16} to the investigation of the effects of communication efficiency and exit capacity on fundamental diagrams for pedestrian moving through an obscure tunnel.
The authors of \cite{lovreglio16} focus on measuring the toxicity due to the smoke inhaled by the occupants.
Using a lattice model, \cite{aube04} looks into the effect of using leader-agents in improving evacuations. When locomotion is biased by a reduced visibility, behavioral rules dominate the choices in the pedestrian flow velocity, see e.g. \cite{pelechano06} for a social force model based on behavioral rules.
Further relevant literature can be found in \cite{tabak10}, \cite{kobes10}, \cite{kobes101}, \cite{jonsson17} and references cited therein.

Despite the obvious influence of distance and vision on social interactions, constraints imposed by the built environment are significant especially when one wants to forecast specific types of collective behaviors of humans during an evacuation in the presence of an unexpected fire (e.g. group formation or boundary-following lanes). Understanding the collective behavior of humans in built environments will certainly lead to a better foundation of environmental psychology as well as to an improved efficiency of way-signaling of route choices, setting the stage for an intelligent crowd management system; see \cite{sorqvist16} for a list of challenges that environmental psychology currently encounters and \cite{becker12} about shortcomings in existing crowd management policies.

All these modeling descriptions reflect the evacuation situation of a given length scale, either {\em micro} (at the pedestrian level) or {\em macro} (at the level of a continuum crowd), or perhaps at some intermediate mesoscopic level (compare \cite{degond13}, e.g.).
In \cite{duong17}, the authors start thinking of connecting relevant microscopic and macroscopic crowd scales by looking for mathematical arguments to ensure the consistency of eventual {\em micro-to-macro} transitions between agent-based and continuum representations.
On the other hand, Richardson in \cite{richardson162} proposes detailed explanations of a {\em micro-macro} model (originally proposed to the computer vision community by Narain {\em et al.} \cite{narain09}) and checks its suitability in a number of test cases (including also a music festival setup). In the framework of this paper, we develop further the {\em micro-macro} approach from \cite{richardson162} and expand it to allow for the presence of a fire as well as for smoke production and dissipation in two different types of confinements. 

The proposed model is hybrid. This is not only because it mixes information captured at two separated space scales (micro and macro), but also because it combines continuum, discrete and stochastic elements; see Section \ref{sec:model} for a detailed presentation of the main model ingredients.

To get an intuitive level of description of the multiscale features of our model, we suggest the following {\em Gedankenexperiment}: Imagine a flow of a fluid, whose internal microstructures (particles) have their own dynamics. Depending on the time-space distribution of particles (i.e. of the approximated density) within the chosen geometry mimicking the built environment, the fluid exhibits both compressible and incompressible regions with {\em a priori} unknown locations. The model is able to automatically detect these zones, by relying on a switching on/off relation, usually referred in the literature as unilateral incompressibility constraint. Interestingly, since in our setting we do not allow for agglomerations of particles (groups) exceeding a certain threshold ({\em a priori} prescribed maximum-allowed density), a macroscopic repulsion term arises in the setting of the equations for the microscopic dynamics (cf. Section \ref{sec:agent}). Our simulations show pressure plots that clearly exhibit the effect the macroscopic repulsion has on the local arrangements of individual evacuees in the neighborhood of densely packed zones. We then point out the effect the environment knowledge has on our evacuation scenario, where both fire and smoke are present.
While contributions exist on modelling the mixing of pedestrians with different levels of knowledge in evacuation scenarios (see e.g. \cite{tan15} for environment knowledge modelling in a lattice model), to our knowledge, there exists no contribution that combines these crowd dynamics aspects with the appearance and evolution of both smoke and fire, even though several contributions have indicated its significance on human behaviour (see e.g. \cite{bryan99},\cite{horiuchi86})

The paper is organized as follows. Section \ref{sec:model} contains the presentation of our multiscale crowd model. The obtained simulation results are reported in Section \ref{sec:results}. We conclude the paper with a short discussion of the main outlets as well as with a concrete to-do-list for the upcoming work. Details on the implementation of our model and related variants can be found in \cite{mercurial}.

\section{Multiscale crowd simulation framework}
\label{sec:model}
We propose a micro-macro pedestrian dynamics model composed of a space-continuous agent-based representation (the micro level) and a continuum compressible flow model (the macro level). Hybrid crowd models, in varying degree resembling ours, have been presented, for instance, in \cite{borve15} and \cite{qi16}. 

The main ingredients of our micro-macro model are:
\begin{itemize}
\item associated evacuation scenario, see Section \ref{sec:evac};
\item prototypical building geometry, see Section \ref{sec:geo};
\item fire model, see Section \ref{sec:fire}; 
\item smoke  model, see Section \ref{sec:smoke};
\item micro-macro evolution equations for the dynamics of the crowd, see Section \ref{sec:path}, Section \ref{sec:agent}, and Section \ref{sec:continuum}.
\end{itemize}

\section{Evacuation scenario}\label{sec:evac}

As case study, we consider the evacuation of a single floor in a large enclosure with interior walls and multiple exits much like a large office building or shopping mall, filled with occupants. In this enclosure a hydrocarbon pool fire is ignited. The model investigates the situation where all occupants have acknowledged the need to evacuate, although they are not necessary aware where the fire is located. While evacuating, the radiation and heat from the fire repels occupants from moving too close to the location of the fire. A particularly important aspect of our investigation is that the fire produces a dense smoke that reduces the visibility of the occupants, diminishing the locomotion speed.

The occupants belong two one of two groups: those familiar with the local geometry, and those who have to rely on visual cues and on following other people.
The complexity of the geometry accentuates the relevance of the environment knowledge on the evacuation.

\section{Geometry}\label{sec:geo}

\begin{figure}[ht]
    \centering
    \includegraphics[width=0.5\textwidth]{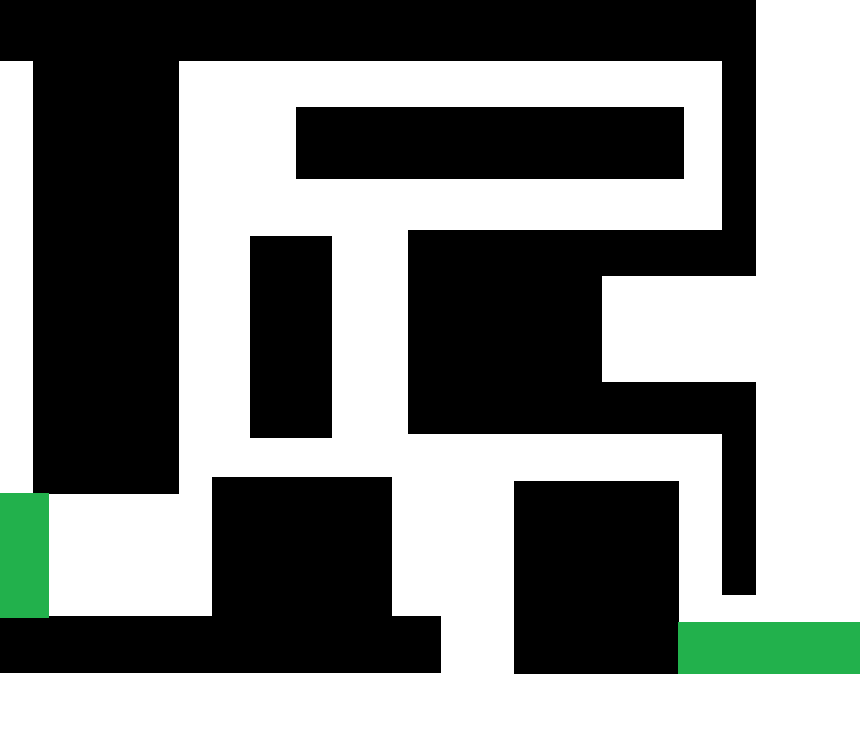}
    \caption{The geometry in our case study. Obstacles are colored black, the two exits are colored green.}
    \label{fig:example}
\end{figure}

Our simulation takes place in a two-dimensional rectangular domain, which we refer to as $\Omega$. 
Parts of the domain are filled with rectangular obstacles; their collection is denoted by $G$. Multiple exits are available. 
At the start of the evacuation, $N$ evacuees are positioned  inside $\Omega$. We assume all evacuees have acknowledged the need of evacuation and are attempting to move towards the nearest exit.
Figure~\ref{fig:example} shows an example geometry. Its complex structure asserts the need of environment knowledge to find  the exit in a reasonable time frame. As a specific feature of our choice of geometry, when a fire blocks one of the corridors, alternative options to the exit are available from each location.

\subsection{Design fire}\label{sec:fire}

The evolution of a fire can be classified in several stages, encompassing the growth stage, the fully developed stage (post-flashover stage), and the decay stage (\cite{drysdale2011}).
In this model, our design fire takes the form of a cylinder-shaped pool fire based on a hydrocarbon fuel that has fully developed and retains a constant heat release throughout the simulation.
Experiments (cf. e.g. \cite{bystrom2017}) show that due to limited air supply in enclosed environments, fires can retain in this stage for a long time.

To describe an experimentally viable scenario, we use measurements from the experiments reported in \cite{smith1996}, \cite{hirschler91}, and \cite{bystrom2017} and fire physics relations explained based on basic thermodynamics in \cite{cengel97}.
Let $\vec{x}_0$ denote the center of the fire that has radius $r_0$. The temperature $T_F$ of our design fire as a function of heat release rate $R$ is given by the following proportionality relation
$
    R \approx h_c ( T(\vec{x})-T_0)= T_F\exp\left(-\kappa\frac{|\vec{x} - \vec{x}_0|}{L}\right),
$
where heat transfer coefficient $h_c$ is chosen to be $300 \si{\watt\per\kelvin}$, $\kappa$ represents the convection heat transfer coefficient, chosen to have a typical value of $20 \si{\watt\per\meter\squared\per\kelvin}$, $T(\vec{x})$ approximates a stationary temperature distribution within the geometry, whose typical length is $L$, and $T_0$ denotes room temperature.
As an indication, \cite{smith1996} shows heat release rates for fully developed hydrocarbon fires to be in the range of 2.5 and 4.5 \si{\kilo\watt}. Accordingly, we  choose $R=3 \si{\kilo\watt}$, yielding a fire temperature of $T_F = 1293 \si{\kelvin}$. This calculation assumes a room temperature of $293 \si{\kelvin}$, but notice that the applicability of the model is not limited to these values.


In Section~\ref{sec:path}, we elaborate on the effects of fire on the choice of evacuations paths.

\subsection{Smoke production}
\label{sec:smoke}

We model the production and spreading of smoke as a diffusive-dominated process. While this does not capture the plume dynamics as a CFD model might, our goal is to relate the smoke density to the visual acuity of the occupants, instead of recovering the exact space-time dynamics of the smoke.
The experiments performed in \cite{tewarson08} show a linear relationship between the smoke emission rate and the heat release rate. We adopt this relationship and take $y_s=0.07 \si{\gram\per\kilo\watt\per\second}$ as a production coefficient.

The smoke density $s(\vec{x},t)$, measured in grams per cubic meter, is assumed to respect a diffusion-drift-reaction equation of type
\begin{equation}
\displaystyle
\begin{cases}
    \partial_t s = \operatorname{div}(D\nabla s) - \operatorname{div}(\vec{v}s) + y_s H(\vec{x})&\mbox{ in } \Omega \setminus G,\\
    (-D\nabla  s+ \vec{v}s)\cdot \vec{n} = 0 &\mbox{ on } \partial \Omega \cup \partial G,\\
    s(\vec{x},0) = 0 &\mbox{ in } \Omega,
\end{cases}
\label{eq:smoke}
\end{equation}
where $D$ represents the smoke diffusivity determined by the environment, $\vec{n}$ is the outer normal vector to $\partial \Omega \cup \partial G$, $\mathbf{v}$ is a given drift corresponding to, for instance, ventilation systems or indoor airflow,  while $H(\vec{x})$ encodes the shape and intensity of the fire, viz.

\begin{figure}[ht]
    \centering
    \includegraphics[width=0.6\textwidth]{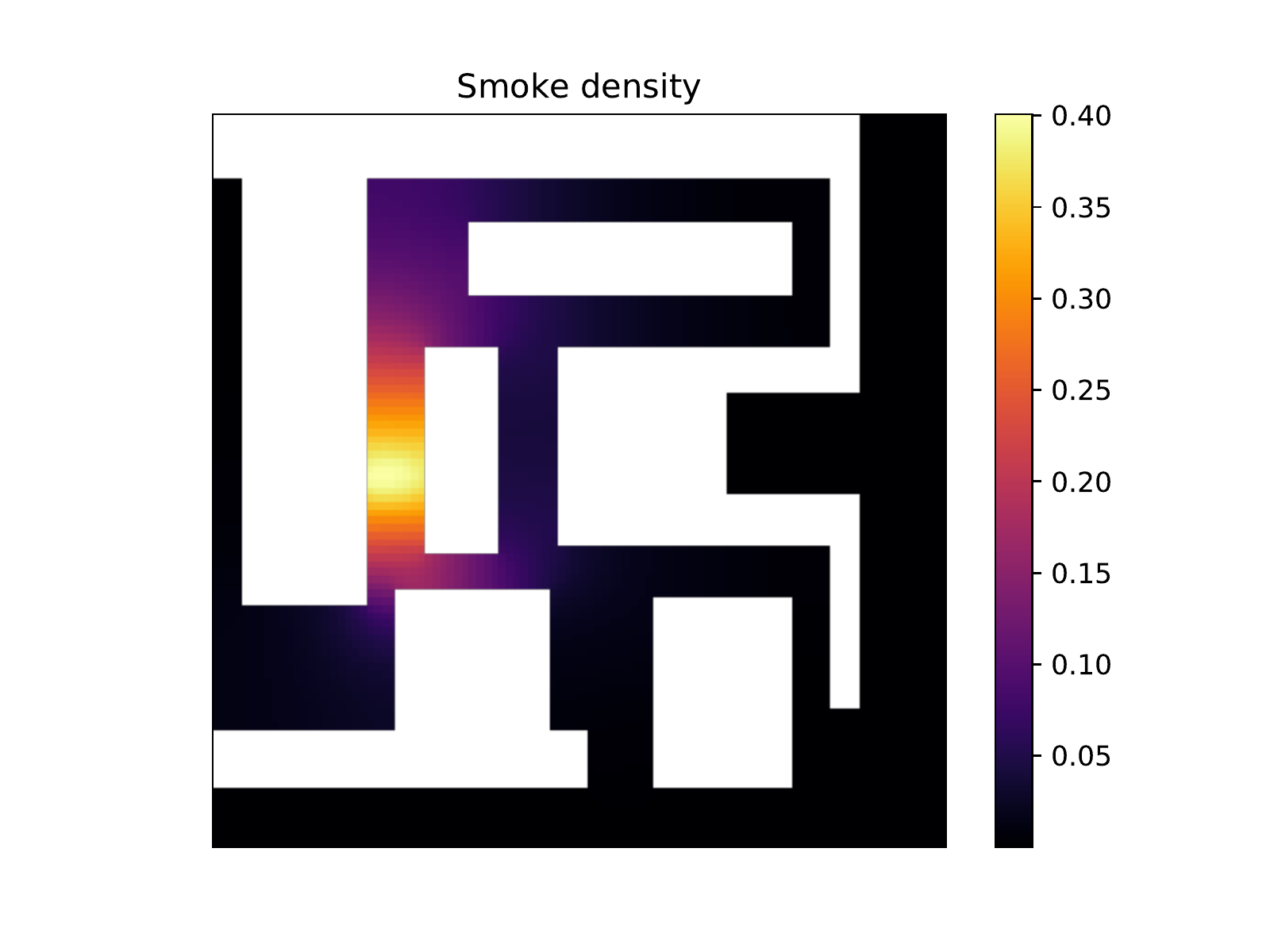}
    \caption{Smoke density in the environment at $t=60$ in grams per cubic meter.}
    \label{fig:smoke_prop}
\end{figure}

A snapshot of the smoke density in the geometry under consideration is presented in Figure~\ref{fig:smoke_prop}.
\begin{equation}
    H(\vec{x}) = \begin{cases}
        R &\mbox{ if }|\vec{x} - \vec{x}_0| < r_0\\
        0 &\mbox{ otherwise }
    \end{cases}.
    \label{eq:fire}
\end{equation}

\subsection{Sight extinction and speed decrease}

The smoke density diminishes the sight radius of the occupants. According to the measurements reported in \cite{mulholland00}, a typical extinction coefficient for a hydrocarbon fuel is $\sigma = \SI{10}{\meter\squared\per\gram}$.
This leads to the following empirical relation between the sight radius $r$ and the local smoke density $s(\vec{x},t)$, taken from \cite{jin97}:
\begin{equation}
    r(s(\vec{x},t)) = 3/(\sigma s(\vec{x},t)) \mbox{ for } x\in \Omega \setminus G, \ t\geq 0.
    \label{eq:smoke_sight_relation}
\end{equation}
\cite{jin97} describes an experimentally acquired relation between the smoke extinction and the walking speed. This relation is expressed here as
\begin{equation}
    v_s(\vec{x},t) = 1.1 - 0.9\sigma s(\vec{x},t),
    \label{eq:smoke_speed_relation}
\end{equation}
where $v_s$ does not take values outside the range of $1.1$ and $0.2 \si{\meter\per\second}$.
 
\subsection{Pedestrian dynamics}
\label{sec:path}

Our agent-based representation refers to two groups of evacuees, one familiar and one unfamiliar with the building geometry. These two groups will be referred to as \emph{residents} and \emph{visitors}, respectively. The dynamics of the residents are governed by first-order differential equations encoding optimal environment knowledge, while the dynamics of the visitors are governed by second-order differential equation based on flocking behaviour. 
Note that the difference in model order between the two evacuee groups reflects the fact that second group of evacuees  meets a large inertia when taking decision on route choices. While the evacuees governed by the first order model are able to change direction instantaneously, the evacuees governed by the second order model in \eqref{eq:x_b} have their velocity adjusted through the application of (social) forces, reflecting their longer reaction time mainly due to the unfamiliarity of the  environment.

Interaction between occupants is transferred to a macroscopic description level by using a continuum flow model.
In this continuum model, agents are evaluated as a crowd density, moving according to a flow velocity field that ensures that the crowd satisfies certain density constraints.
The micro and macro model representations are coupled and communicate in each time step.

The continuum representation allows us to model  efficiently the interactions within the crowd in high-density regions, while the agent-based representation gives control over the individual trajectories. Yet by avoiding computing interactions at agent-based level, we are able to increase  significantly the speed of the simulation.

In addition to the effects of smoke described in the previous section, the fire impacts the trajectories as follows. The elevated temperature near the fire repels the evacuees: residents close to the fire will take notice and find an alternative route to an exit, while visitors are forced to simply move in an opposite direction. 
These dynamics are described in the next section.

\section{Agent-based dynamics}
\label{sec:agent}

Residents and visitors are seen as elements from the sets $X_A = \{a_1,...,a_{N_A}\}$ and $X_B = \{b_1,...,b_{N_B}\}$, respectively, where $N_A + N_B=N$. For ease of notation, we denote the complete set of evacuees with $X:=X_A \cup X_B$.
Since we want to emphasize the difference in environment knowledge between the evacuee groups, we employ different path finding algorithms for each group.

\subsection{Residents}
To describe the motion of the residents we use a potential field model proposed by Hughes in \cite{hughes02} and adapted in \cite{treuille06}. It functions similarly to a floor field function, its counterpart in lattice models like \cite{tan15} and \cite{cao2014}. We augment the potential field model to account for the presence of obstacles, fire and smoke.

The potential field is based on the ``minimization of effort'' principle, serving as a dynamic generalized distance transform. For each point $\vec{x}$ in the domain, we introduce a \emph{marginal cost field} $u(\vec{x})>0$, defined as
\[ u(\vec{x}) = \alpha + u_{\mathrm{obs}}(\vec{x}) + wH(\vec{x}).\]
The marginal cost field is built up from a base level of constant walking effort $\alpha>0$, information on the geometry and the obstacles $u_{\mathrm{obs}}$, and information on the fire source $w$. Here, $w$ takes value 1 if the resident is aware of the location of the fire and 0 otherwise.

Let $S$ be a path from point $\vec{x}_p$ to point $\vec{x}_q$. Then the effort of walking on the path $S$ can be expressed as
\[\int_S u(\vec{\xi})d\vec{\xi}=\int_S\alpha + u_{\mathrm{obs}}(\vec{\xi}) + wH(\vec{\xi})d\vec{\xi}.\]
At the beginning of the evacuation, $w$ is 0 for all residents. When a resident experiences a significant increase in temperature because of his proximity to the location of the fire, $w$ is set to 4 and $S$ changes, and as a result, the fire is avoided.

Recall that $G\subset \Omega$ is the set of all inaccessible locations in the geometry (i.e. those parts of $\Omega$ covered by obstacles). Then for all $\vec{x}\in \Omega$, the geometry information (i.e. the obstacle cost field) can be expressed as
\begin{equation}
    u_{\mathrm{obs}}(\vec{x}) =
    \begin{cases}
        \infty &\mbox{if }\vec{x} \in G,\\
        \frac{1}{|d(\vec{x},G)|} & \mbox{if }\vec{x} \notin G\mbox{ and }d(\vec{x},G) \leq r_G,\\
        0 &\mbox{if } d(\vec{x},G) > r_G,
    \end{cases}
    \label{eq:pot_obs}
\end{equation}
where $r_G$ is a small parameter roughly of the order of the averaged size of the evacuees. The obstacle cost makes sure that obstacle locations are inaccessible, and it adds a tiny layer of repulsion around each obstacle to ensure the basic fact that evacuees do not run into walls.

The preferred path $S^*$ for an evacuee with location $\vec{x}_p$ and motion target $\vec{x}_q$ is determined as
\begin{equation*}
    S^* = \argmin_S \int_S u(\vec{\xi})d\vec{\xi},
\end{equation*}
where we minimize over the set of all possible motion paths $S$ from $\vec{x}_p$ to $\vec{x}_q$.
In the simulation, the evacuees from $X_A$  are aware of all exits, and the optimal path $S^*$ is made available by means of a potential function $\Phi$, a solution to the equation
\begin{equation}
    \left|\left|\nabla \Phi(\vec{x})\right|\right| = u(\vec{x}),
    \label{eq:eikonal}
\end{equation}
where $||\cdot||$ denotes the standard Euclidean norm. Visitors do not have access to the optimal paths.

\begin{figure}[ht]
\centering
\begin{minipage}{.5\textwidth}
        \centering
    \includegraphics[width=\textwidth]{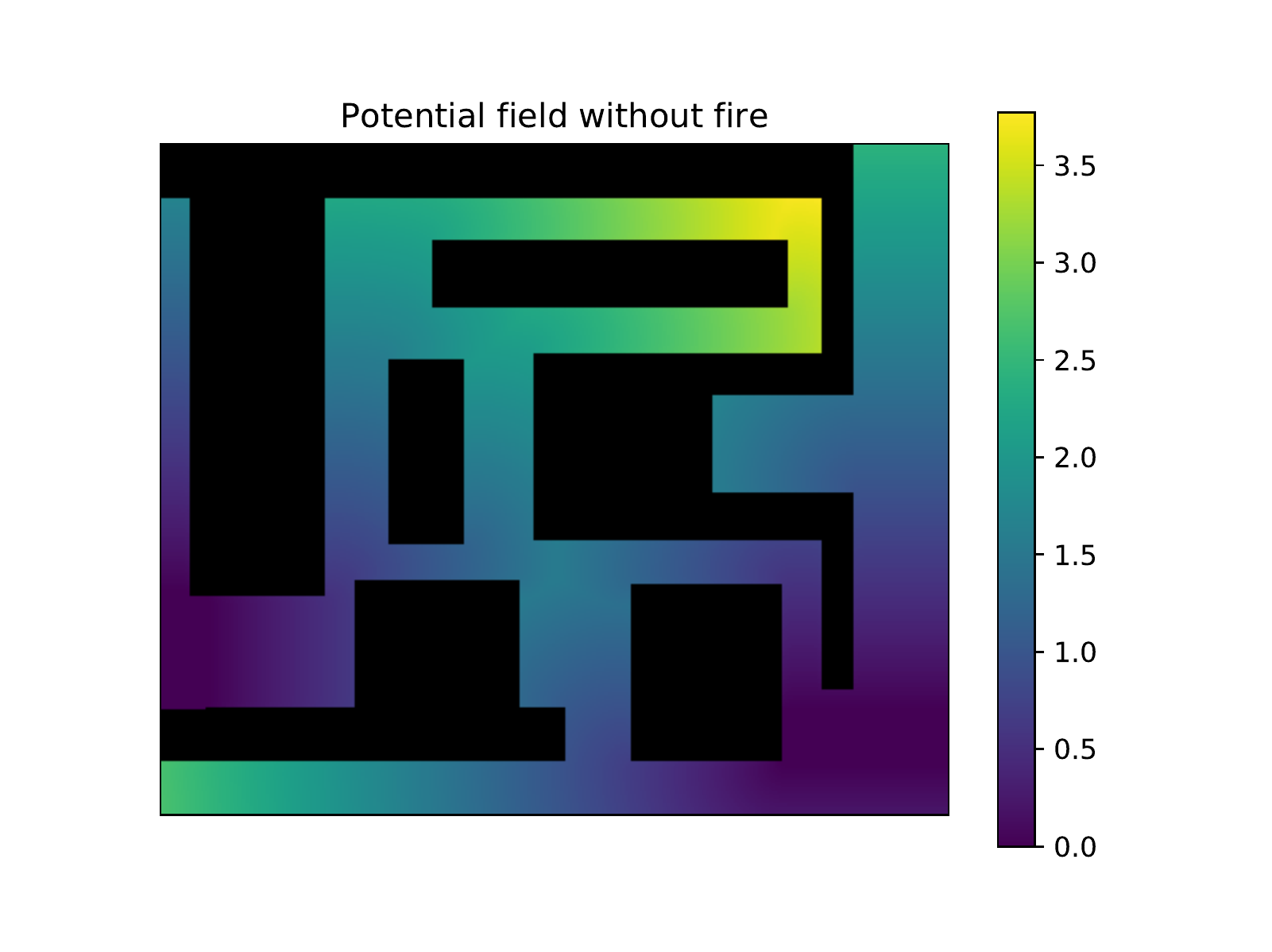}
    \caption{Potential field $\Phi$ for the environment of the case study, not taking any fire into account.}
    \label{fig:potential_standard}
\end{minipage}%
\hfill
\begin{minipage}{.4\textwidth}
        \centering
        \includegraphics[width=\textwidth]{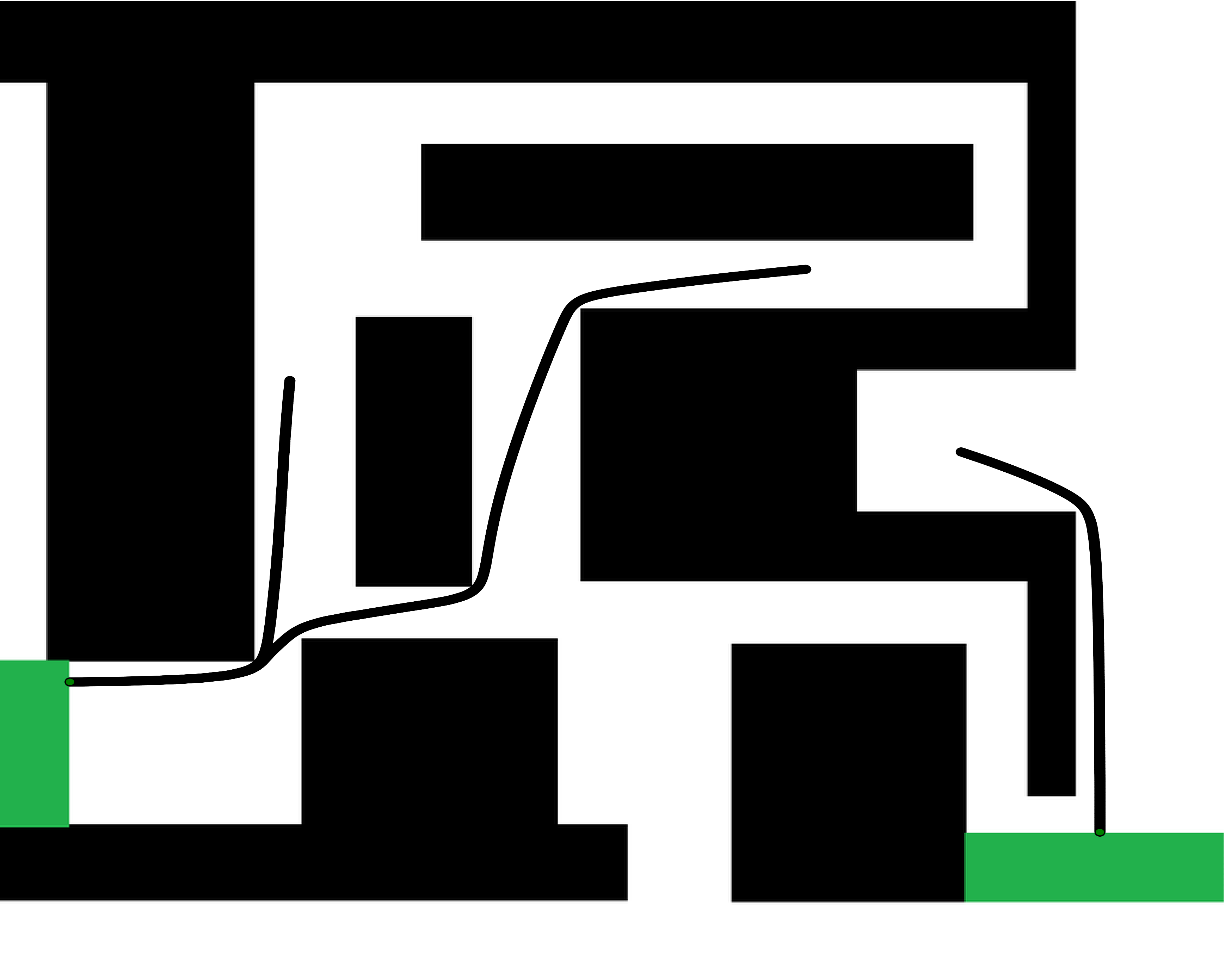}
        \caption{Paths generated from the potential field in Figure~\ref{fig:potential_standard}.}
        \label{fig:evac_path_standard}
\end{minipage}
\end{figure}

\begin{figure}[ht]
\centering
\begin{minipage}{.5\textwidth}
        \centering
    \includegraphics[width=\textwidth]{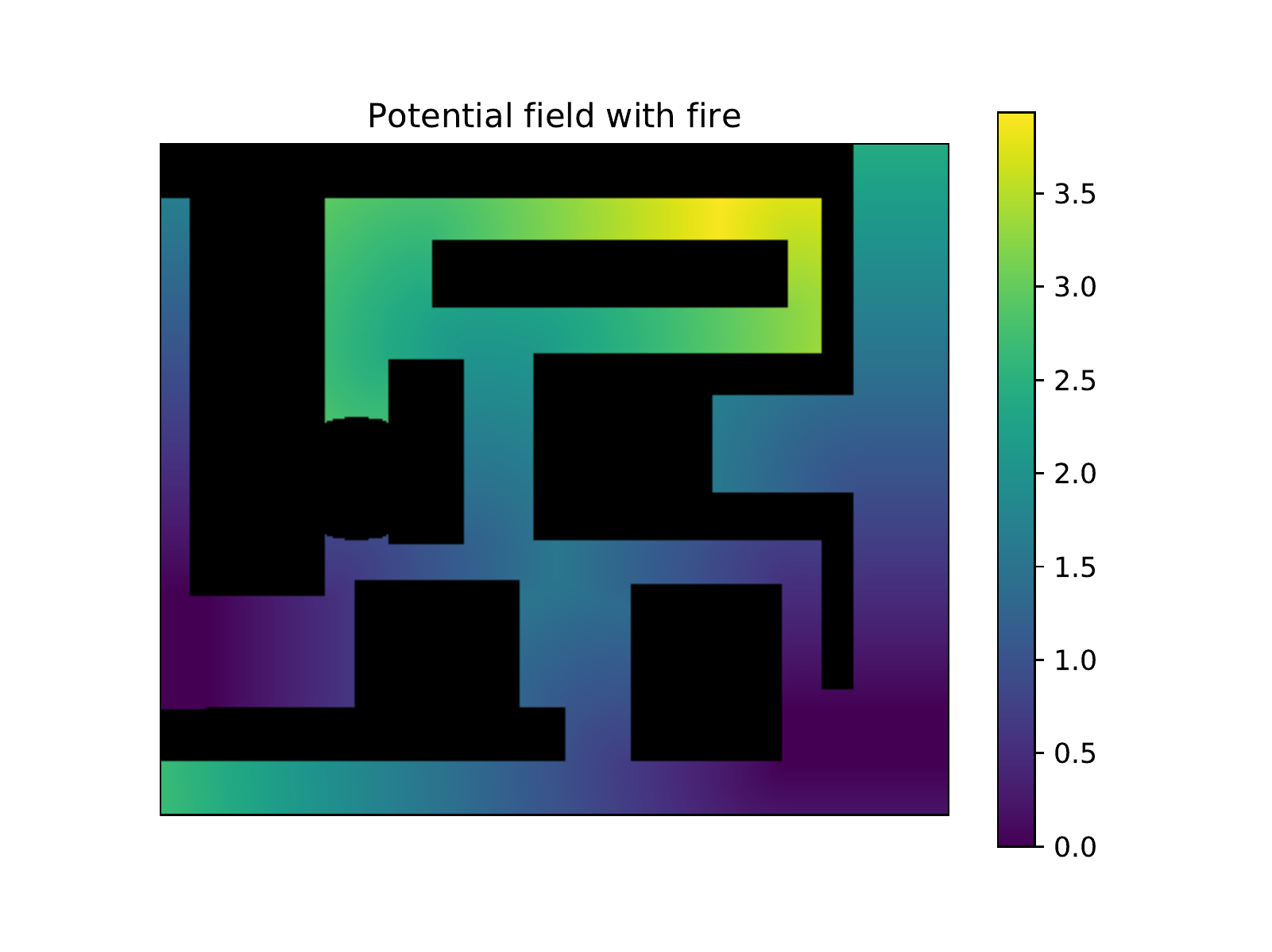}
    \caption{Potential field $\Phi$ for the environment of the case study aware of the fire location.}
    \label{fig:potential_fire}
\end{minipage}%
\hfill
\begin{minipage}{.4\textwidth}
        \centering
        \includegraphics[width=\textwidth]{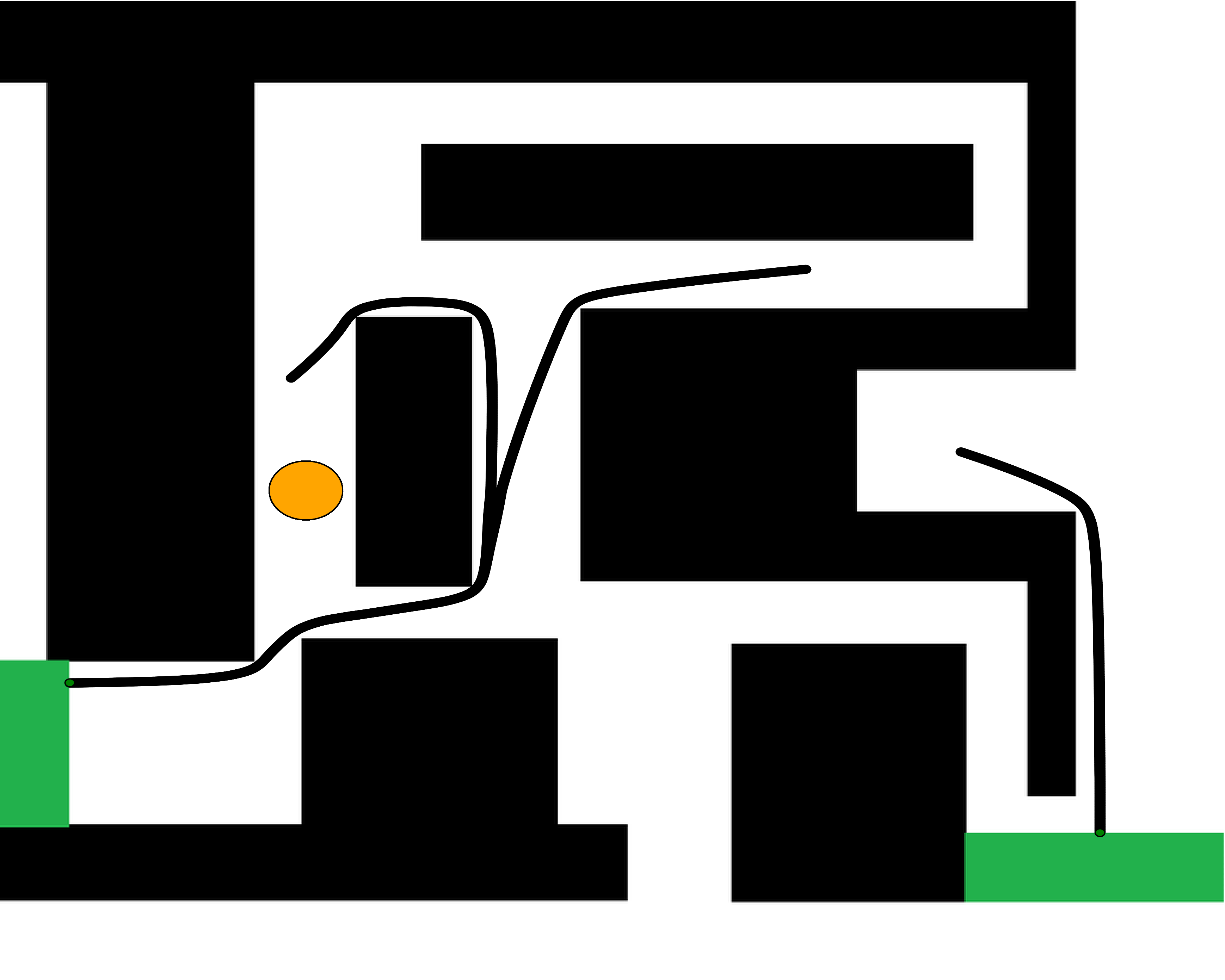}
        \caption{Paths generated from the potential field in Figure~\ref{fig:potential_fire}, avoiding the fire.}
        \label{fig:evac_path_fire}
\end{minipage}
\end{figure}

Figure~\ref{fig:potential_standard} and Figure~\ref{fig:evac_path_standard} display the potential field and the corresponding paths for our case study.
Figure~\ref{fig:potential_fire} and Figure~\ref{fig:evac_path_fire} display the adaption for residents that are aware of the fire locations and take an alternative route out.

Let $\vec{x}_{a_i}(t)$ denote the position of resident $a_i$ at time $t$. The dynamical system that governs his motion is expressed as:
\begin{equation}
\begin{cases}
    \displaystyle
        \frac{d\vec{x}_{a_i}}{dt} &= -v_s(\vec{x}_{a_i},t)\frac{\nabla\Phi(\vec{x}_{a_i}) - \nabla p(\vec{x}_{a_i},t)}{||\nabla\Phi(\vec{x}_{a_i}) - \nabla p(\vec{x}_{a_i},t)||},\\
        \vec{x}_{a_i}(0) &= \vec{x}_{a_i,0},
\end{cases}
\label{eq:x_a}
\end{equation}
where $\vec{x}_{a_i,0}$ represents the initial configuration of the evacuees and $v_s$ represents the evacuation speed defined in \eqref{eq:smoke_speed_relation}.
In \eqref{eq:x_a}, $p$ is a macroscopic pressure term that influences pedestrian interactions at the macroscopic scale. We postpone its definition to Section~\ref{sec:continuum}.

\subsection{Visitors}
Since visitors are unfamiliar with their environment, they rely solely on information from their fellows. This is a modelling assumption based on experimentally confirmed behavior in primates \cite{meunier06} and which has been applied in other evacuation models as well (e.g. \cite{helbing00}). To represent this, we choose to apply a Cucker-Smale-like model which averages the velocity of nearby evacuees (introduced in \cite{cucker07}). This swarming model is adapted with a Brownian term $\mathbf{B}_i$ representing disorienting and chaotic effects of an evacuation in an unknown environment, and a repulsion from the fire source $\nabla H$. We note that according to the definition in \eqref{eq:fire}, $H$ is not differentiable. To obtain this gradient, we smoothen $H$ with a mollifier.

We denote the position and velocities from evacuee $i$ of $X_B$ as $\vec{x}_{b_i}$ and $\vec{v}_{b_i}$, and positions and velocities from member $j$ of the complete evacuation set $X$ as $\vec{x}_j$ and $\vec{v}_j$ respectively.
We express the motion of visitor $b_i$ as
\begin{equation}
\begin{cases}
\displaystyle
    \frac{d\vec{v}_{b_i}}{dt} &= \sum_{j\in X} (\vec{v}_{j} - \vec{v}_{b_i})w_{ij}- \nabla {H}(\vec{x}_{b_i},t)+ \mathbf{B}_i(t)\\
     &+ \frac{\vec{v}_{b_i} - \nabla p}{||\vec{v}_{b_i} - \nabla p||}v_s\left(\vec{x}_{b_i},t\right),\\
     \frac{d\vec{x}_{b_i}}{dt} &= \vec{v}_{b_i},\\
    \vec{v}_{b_i}(0) &= \vec{v}_{b_i,0},\\
    \vec{x}_{b_i}(0) &= \vec{x}_{b_i,0}.
\end{cases}
\label{eq:x_b}
\end{equation}
In \eqref{eq:x_b} $w_{ij}$ are weight factors, decreasing as a function of distance, defined as
\begin{equation}
    w_{ij}(s) \sim \frac{1}{r_s^2}\exp\left(-\frac{|\vec{x_{b_i}} - \vec{x_{j}}|^2}{r_s^2}\right).
    \label{eq:weight_factor}
\end{equation}
In~\eqref{eq:weight_factor}, $r_s$ is the sight radius (recall \eqref{eq:smoke_sight_relation}) in the evacuees location.
It should be noted that we do not take into account those walls that block the transfer of information between evacuees, since they are ignored in \eqref{eq:weight_factor}. However, in the simulations described in Section~\ref{sec:results}, the size of the walls generally exceeds the interaction radius.

Another important observation is that in this model, visitors do not know which of the other occupants are residents, and which are visitors themselves; they follow others indiscriminately.

\subsection{Continuum model description}
\label{sec:continuum}

The path finding velocities are not yet corrected for the interaction and avoidance tendency among evacuees.
Introducing some level of averaged information, we choose to model interactions at the continuum level only. This way, we can access inherent macroscopic properties like densities and crowd pressures (recognized as measures in e.g. \cite{moussaid11}). 
Additionally, we are interested in the influence of fire on the crowd pressure, which both act on the evacuees as macroscopic 'social' forces.
To adjust the behavior of evacuees in crowded zones,  we apply a unilateral incompressibility constraint (UIC) as proposed in \cite{narain09}. 
First, we interpolate the agent positions to a macroscopic density $\rho$ and velocity $\vec{v}$. Then we introduce a pressure $p$ that forces the macroscopic density to satisfy a prescribed maximum value $\rho_{\max}$.

Let $\rho(\vec{x},t)$ represent the averaged concentration of evacuees in $\vec{x}$ at time $t$, the measure of crowdedness.
The scalar field $\rho$ is obtained by convolving all evacuee positions $\vec{x}_{i}$ pinned as Dirac distributions $\delta_{\vec{x}_{i}}$, with a smooth interpolation kernel $\psi$
\begin{equation*}
    \rho(\vec{x},t) = \sum_{i=1}^N\left( \delta_{\vec{x}_{i}(t)}\ast \psi  \right)(\vec{x}).
\end{equation*}

We obtain a macroscopic velocity in a similar fashion; by convolving all evacuee velocities $\vec{v}_i$ to a velocity field:
\begin{equation*}
    \vec{v}(\vec{x},t) = \frac{\sum_{i=1}^N\vec{v}_{i}(t)\left( \delta_{\vec{x}_{i}(t)}\ast \psi  \right)(\vec{x})}{\rho(\vec{x})}.
\end{equation*}

Finally, we look for a pressure $p(\vec{x},t)$ that satisfies the following equation
\begin{equation}
    \frac{\partial \rho}{\partial t} = -\operatorname{div}\left(\rho(\vec{v}- \nabla p)\right).
    \label{eq:uic}
\end{equation}
subject to the following unilateral incompressibility constraints (UIC) conditions for all $\vec{x}$ and $t$:
\begin{equation}
    \begin{split}
        p(\vec{x},t) \geq 0, \ \ 
        \rho(\vec{x},t) \geq 0, \ \ 
        p(\vec{x},t)(\rho_{\max} - \rho(\vec{x},t)) = 0.
    \end{split}    
    \label{eq:uic_cond}
\end{equation}

\eqref{eq:uic} together with the conditions in \eqref{eq:uic_cond} forces the groups of pedestrians on every patch in $\Omega$ to satisfy a maximum density. As soon as a patch violates maximum density $\rho_{\max}$, a pressure is introduced that steers pedestrians away from the overcrowded location. As a result, the density decreases until it satisfies the maximum density again.
\begin{figure}[ht]
\centering
\begin{minipage}{.45\textwidth}
        \centering
        \includegraphics[width=0.8\textwidth]{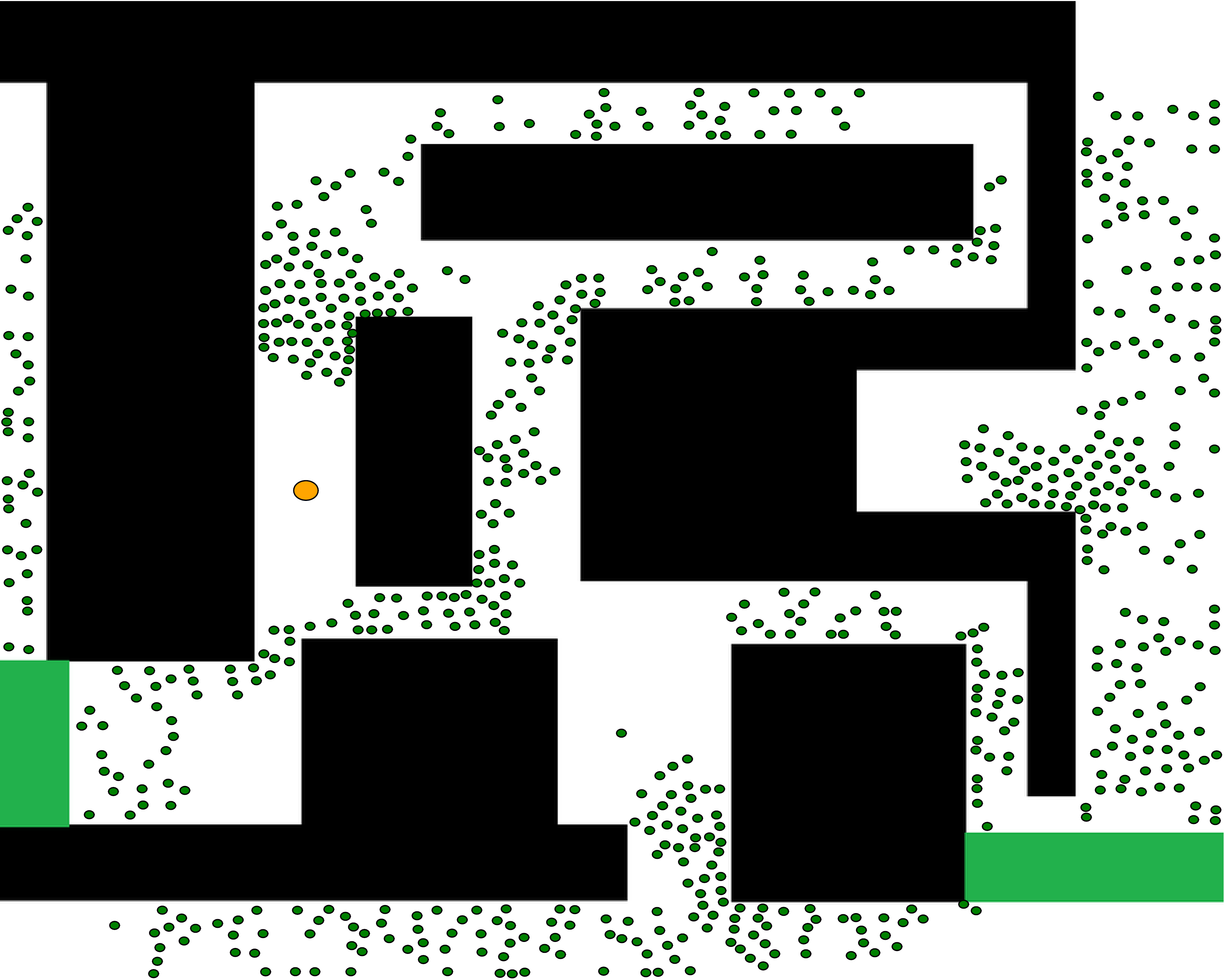}
        \caption{A snapshot of a simulation in the geometry of our case study.}
    \label{fig:snap_pressure}
\end{minipage}%
\hfill
\begin{minipage}{.5\textwidth}
        \centering
        \includegraphics[width=\textwidth]{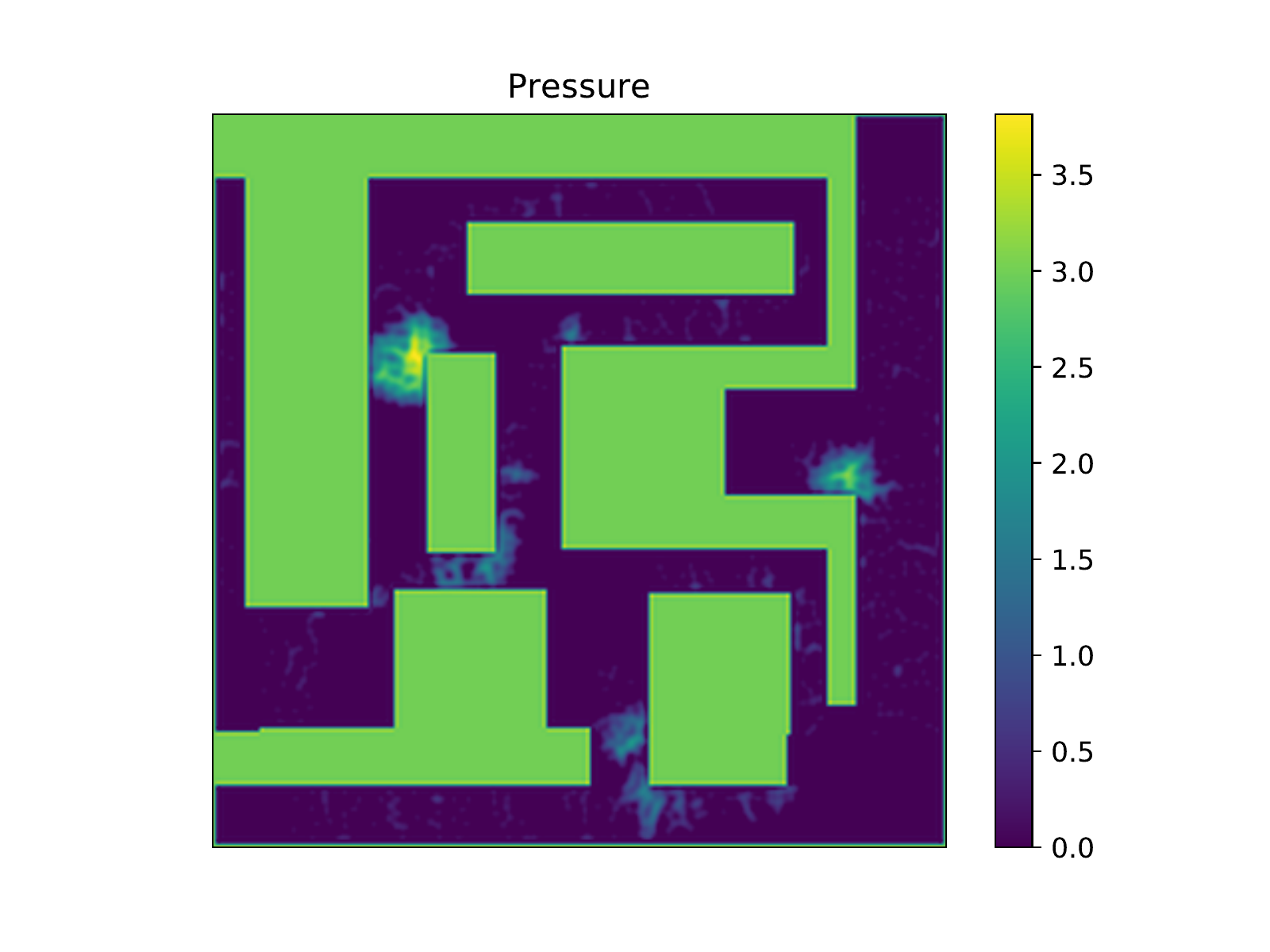}
        \caption{Pressure observed in the simulation snapshot of Figure~\ref{fig:snap_pressure}.}
        \label{fig:pressure}
\end{minipage}
\end{figure}

Figure~\ref{fig:pressure} illustrates a pressure field in an ongoing simulation which contains incompressible zones. The corresponding pedestrian configuration is displayed in Figure~\ref{fig:snap_pressure}.

\section{Discretization}

In our implementation, we discretize $\Omega$ in a rectangular grid fine enough to resolve the obstacles and the exit. We discretize $u$ by averaging \eqref{eq:pot_obs} over each grid cell so that we obtain an approximation of $\Phi$ by numerically solving \eqref{eq:eikonal} using a fast marching algorithm (\cite{tsitsiklis95}). 
This preferred path minimizes the cost of moving from $\vec{x}_p$ to $\vec{x}_q$ where, if this path leads past a fire, a trade-off is made between choosing a short path and maintaining a safe distance from the fire source.

In the pre-processing phase of the simulation, two potential fields are computed, one accounting for the fire, while the other one does not. By keeping track which pedestrians are aware of the fire, the updating of knowledge can be handled in an efficient manner. Equation  
\eqref{eq:uic} is discretized as follows: at each time step of the simulation, in any location $\vec{x}$, if the local density of evacuees $\rho(\vec{x})$ exceeds a threshold $\rho_{\max}$, the UIC takes effect and, if needed, the crowd flow locally changes from a compressible to an incompressible flow regime. We can do this by making sure the macroscopic velocity field $\vec{v}$ is divergence-free in $\vec{x}$.
As a result, $\rho(\vec{x})$ cannot increase and a pressure $p(\vec{x})$ is introduced that exerts a force on the crowd proportionally to $\nabla p(\vec{x})$.
This way, we are able to control the maximum density of the evacuees for any patch in the domain.
The unilateral constraints ensure that no pressure-related forces are applied outside of the high-density zones.

We discretize $\rho$ and $v$ on the same grid as \eqref{eq:smoke} and \eqref{eq:eikonal} as vector objects denoted by $\vec{\rho}^n$ and $\vec{v}^n$, respectively.
The discretized pressure $\vec{p}^n$ is the solution of the quadratic problem:

Minimize $f(\vec{p}^n)$, defined as
\begin{equation}
    f(\vec{p}^n) =  \langle\vec{p}^n, \rho_{\max} - \vec{\rho}^n + \operatorname{div}(\vec{\rho}^n\vec{v}^n - \vec{\rho}^n \nabla \vec{p}^n)\Delta t \rangle,
    \label{eq:discr_pde}
\end{equation}
subject to the constraints 
\[(\vec{p}^n)_i \geq 0, (\vec{\rho}^n)_i \leq \rho_{\max}\mbox{ for all }1\leq i\leq n,\]
where we abuse the notation of $\operatorname{div}\vec{u}$ and $\nabla \vec{u}$ to denote a consistent discrete approximation of continuous counterpart $\operatorname{div} u$ and $\nabla u$, respectively.

Exploiting the complementary relation between $\rho_{\max} - \vec{\rho}^n$ and $\vec{p}^n$, we can convert \eqref{eq:discr_pde} to a linear complementarity problem and solve it using the projected Gauss-Seidel algorithm on each time step without creating a bottleneck for the simulation.
For more details regarding the algorithm and the implementation, we refer to \cite{narain09} and \cite{richardson162} respectively.

Because of our discretization of the UIC, the avoidance tendency is not maintained on distances smaller than the grid size. Therefore, to model size exclusion, we manually need to move evacuees apart to prevent them occupying locations too close to each other.

Each time step, we compute the propagation of the smoke according to \eqref{eq:smoke}, we update the sight radii and the new positions of the agents according to the model in Section~\ref{sec:agent}. When an agent reaches the exit, his residence time is recorded and he is removed from the geometry.
\section{Results}
\label{sec:results}

The results are run in crowd simulation prototyping application \emph{Mercurial} (\cite{mercurial}). This is an open-source framework developed in Python and Fortran to simulate hybrid crowd representations as the one described in Section~\ref{sec:model}. It provides both agent-based- and continuum-level visualizations. \emph{Mercurial} supports the design of arbitrary two-dimensional geometries, and has features for collecting and post-processing simulation results. More details on the structure and implementation of \emph{Mercurial} are found in \cite{richardson162}.

\begin{figure}
    \centering
    \includegraphics[width=0.8\textwidth]{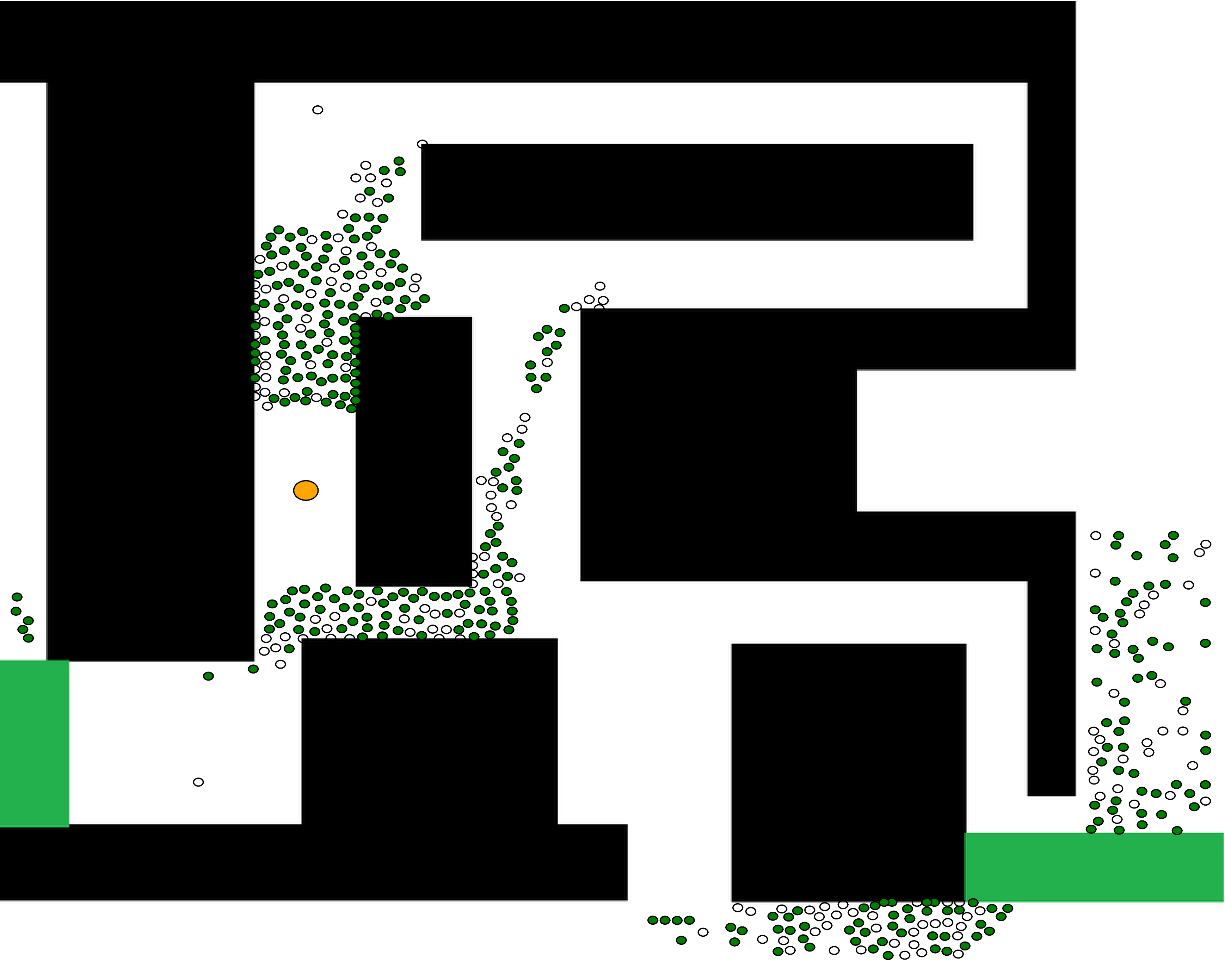}
     \caption{Snapshot after $t=46$ for a visitor ratio of 0.3. Visitors are displayed as empty circles, residents as filled circles. The larger orange circle represents the center of the fire.}
     \label{fig:case05}
\end{figure}
\begin{figure}
    \centering
    \includegraphics[width=0.8\textwidth]{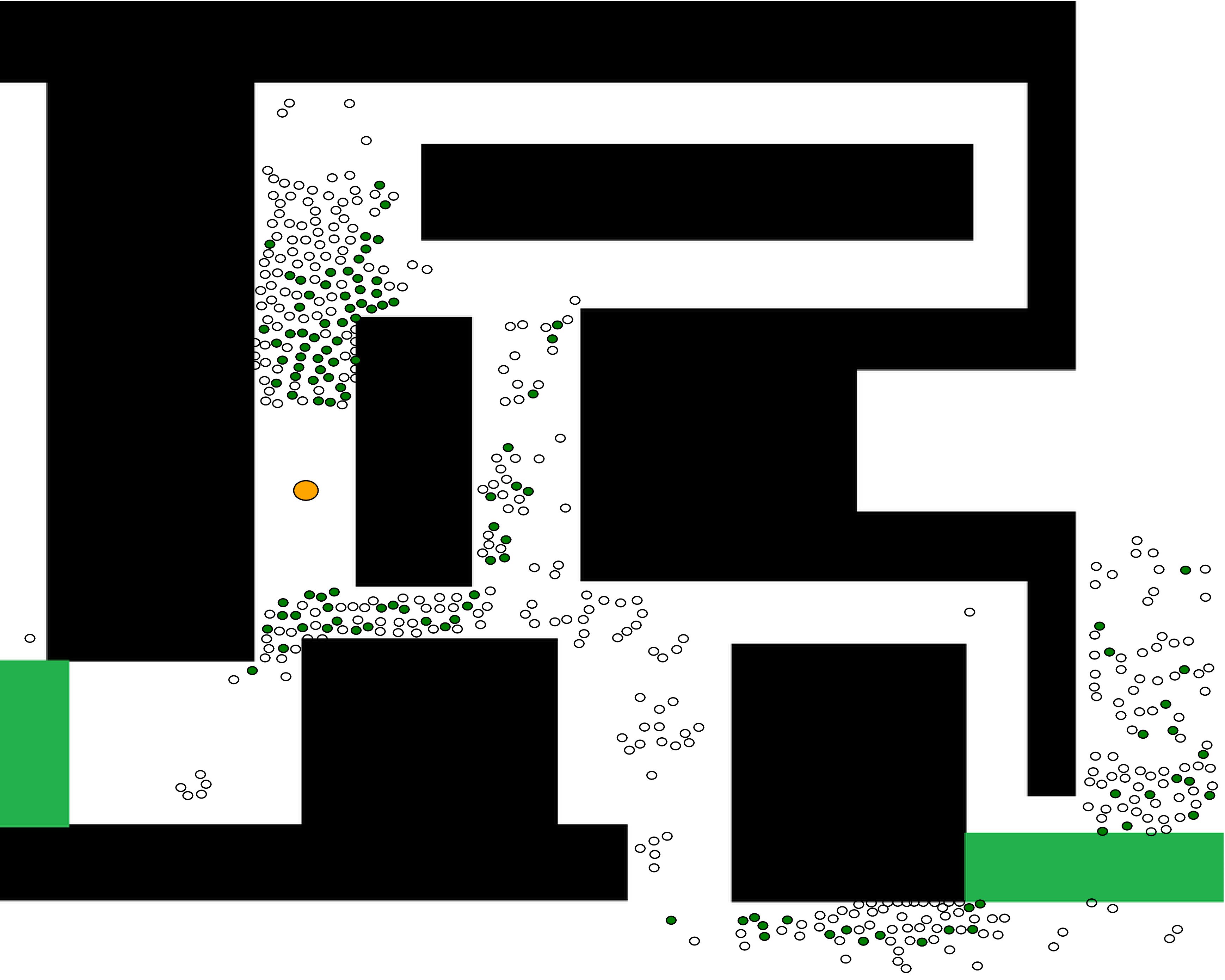}
         \caption{Snapshot after $t=46$ for a visitor ratio of 0.7. Visitors are displayed as empty circles, residents as filled circles. The larger orange circle represents the center of the fire.}
         \label{fig:case95}
\end{figure}

We ran the simulation 15 times in our case study containing 1000 evacuees, each time varying the ratio of visitors and residents. We started with a scenario filled with only residents. We indicate the simulation configuration by the ratio of visitors in the environment ($\frac{N_B}{N_A+N_B}=0$), and increased this number with 0.05 for each subsequent run. Figure~\ref{fig:case05} and Figure~\ref{fig:case95} display two snapshots of the geometry for ratios 0.3 and 0.7, respectively.


We have chosen this environment to resemble the interior of a large building, characterized by corridors. The environment is set up so that without knowledge of the environment, it is possible but quite difficult to find an exit, especially with high densities of smoke.

In Figure~\ref{fig:case05}, we observe a structured evacuation, with almost of the evacuees belonging to a group moving towards the exit. As one would expect, the evacuation in Figure~\ref{fig:case95} progresses more chaotically. Most evacuees move in smaller groups, their direction largely influenced by the outline of the corridor they move in.

\begin{figure}
    \centering
    \includegraphics[width=0.8\textwidth]{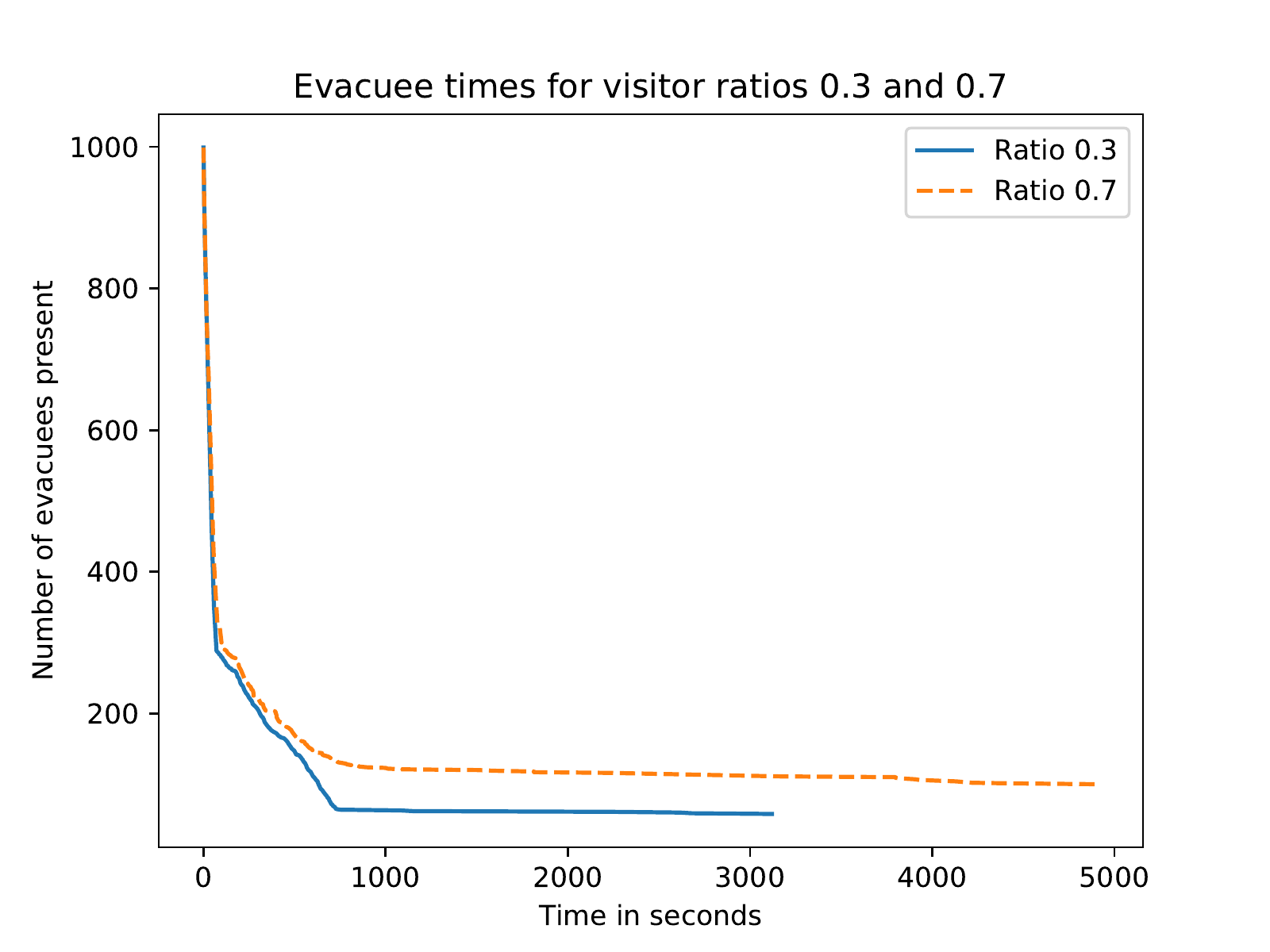}
    \caption{Number of evacuees left at time $t$ for two ratios of occupants.}
    \label{fig:evac_times}
\end{figure}

The larger groups typically are centered around a resident, constituting a ``shepherding effect'': a leader/follower scenario where the leader is surrounded from all sides by a large group of followers. This allows a large number of evacuees to reach the exit in both scenarios, as observed in Figure~\ref{fig:evac_times} which displays the evacuation times for the two cases. This shepherding effect is most strongly present in the beginning of the situation, when the smoke density is still low enough for visitors to maintain a large sight radius. In some configurations, we observed groups of visitor clusters with an $X_A$ leader to move faster than a typical group of residents.

Notice how in both evacuations there exist a large crowded zone close to the fire location. This is due to the lack of information the residents have about the location of the fire. Attempting to take the fastest route to the exit, takes them through hordes of residents and visitors drawn towards the same location. Because of these conflicting directions, the visitors are not able to consistently follow other evacuees and the evacuation stagnates.
This does not happen in the other corridors, regardless of their level of congestion.

Figure~\ref{fig:evac_times} depicts the evacuation process as a function of time. 
In the initial stage of evacuation, egress is controlled by the capacity of the corridors and exits (revealed by the near-constant slope).
After that time, the bulk of evacuees has been removed from the scene and the remainder consists of evacuees with a high initial distance to the nearest exit.
Note that in spite of the large difference in evacuee ratios, the evacuation times for most residents are quite similar.
The observation that large agent systems can be driven with a relatively small number of 'active particles' is consistent with observations made in other contributions on particle systems (cf. e.g. \cite{kusters17}).
This illustrates the effect of the leader-follower dynamics. 
The evacuation stagnates after all residents have left, leaving only the visitors with a lack of information.

\begin{figure}[ht]
\centering
\begin{minipage}{.45\textwidth}
        \centering
    \includegraphics[width=\textwidth]{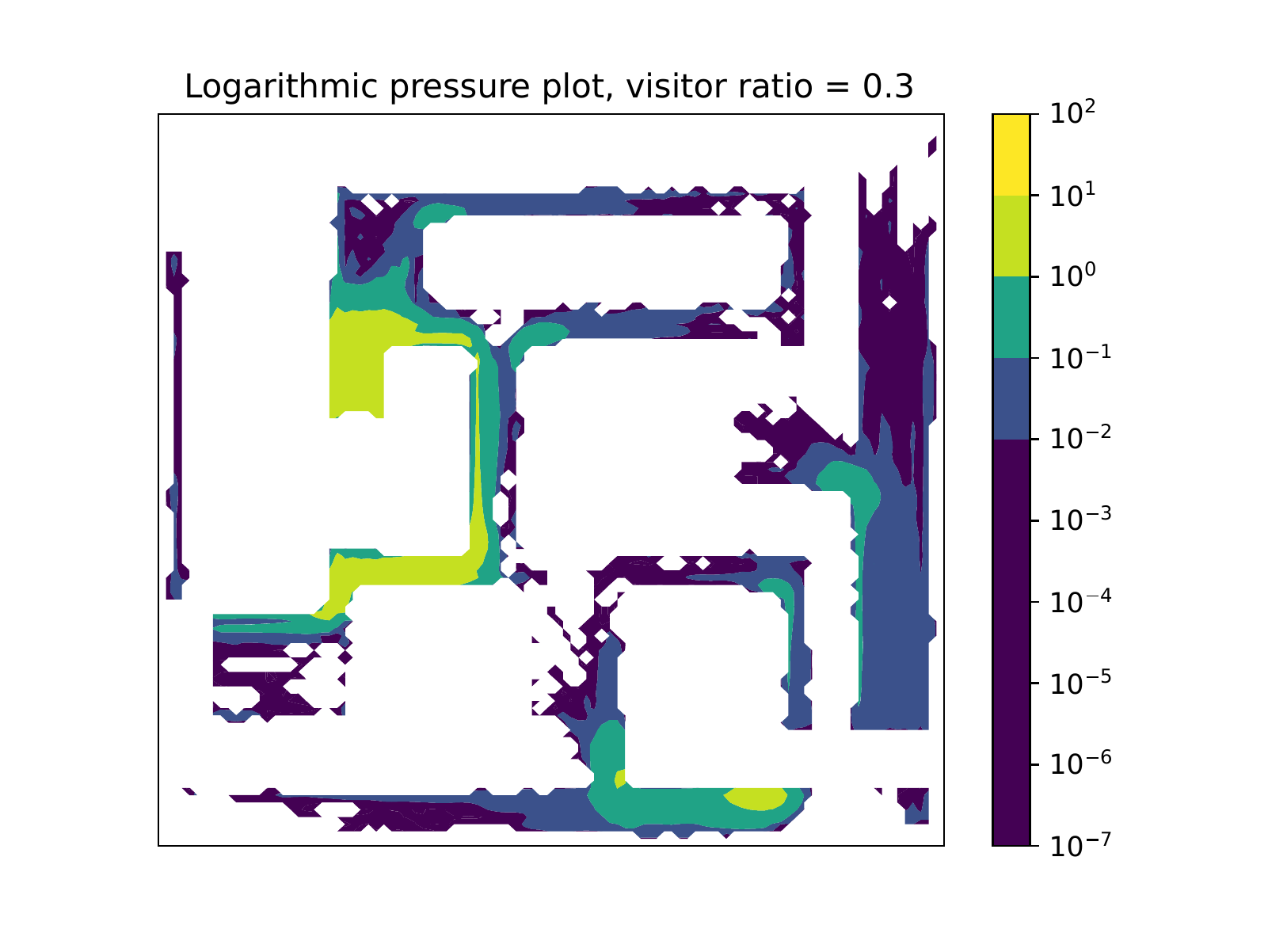}
    \caption{Logarithmic heat-map of pressure zones that developed in the scenario for a visitor ratio of 0.3.}
    \label{fig:pressure_residents}
\end{minipage}%
\hfill
\begin{minipage}{.45\textwidth}
        \centering
        \includegraphics[width=\textwidth]{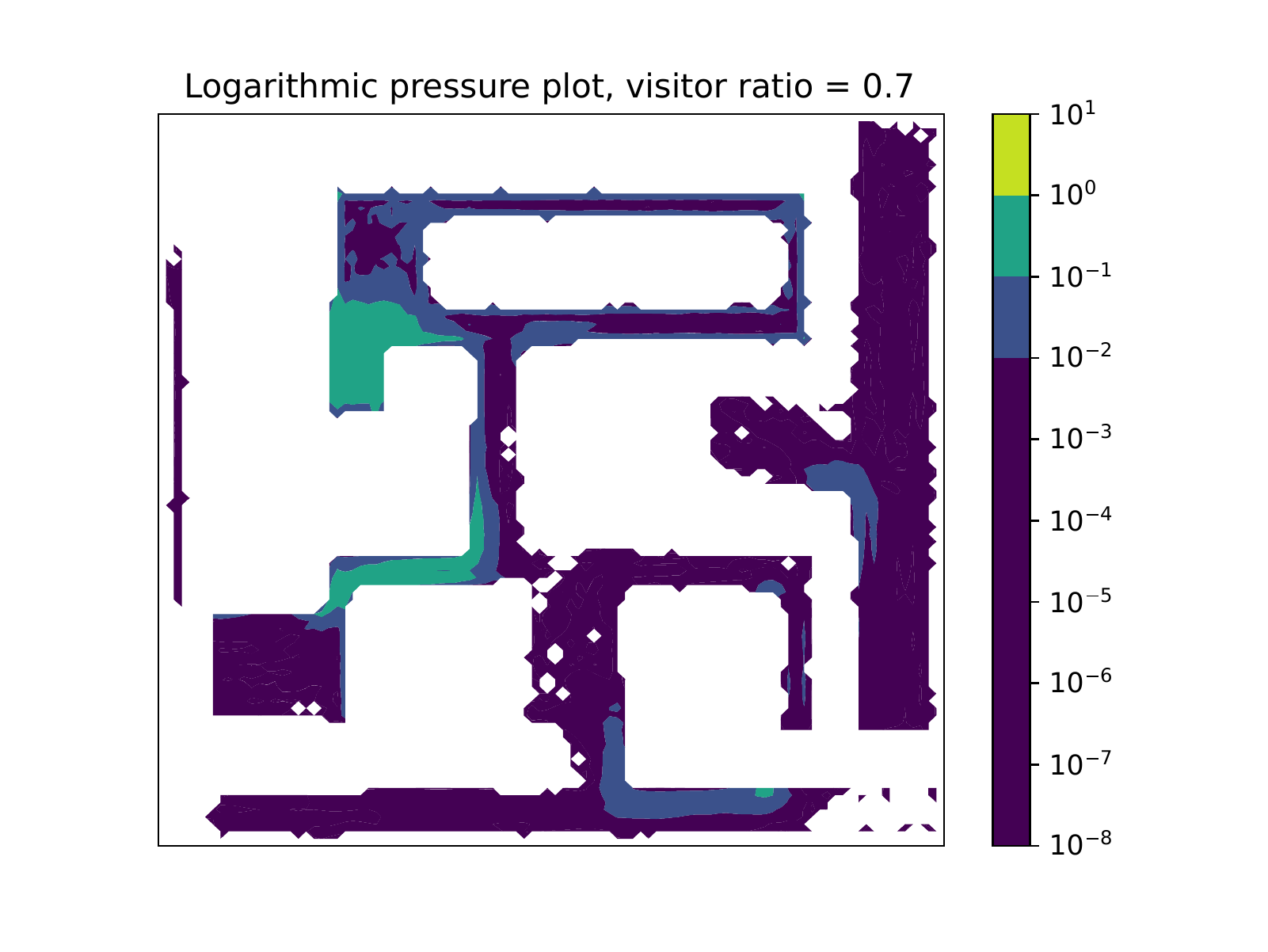}
        \caption{Logarithmic heat-map of pressure zones that developed in the scenario for a visitor ratio of 0.7.}
        \label{fig:pressure_visitors}
\end{minipage}
\end{figure}

The difference in evacuees behavior is also revealed by the formation of the incompressibility zones, shown in Figure~\ref{fig:pressure_residents} and Figure~\ref{fig:pressure_visitors}.

The pressures experienced in simulations with low visitor ratios are substantially higher. This can be attributed to the persistence in direction the residents display. Whenever a corridor appears clogged, the visitors are more prone to find an alternative exit. The difference in the pressure plots indicates that the visitors are more successful in avoiding the fire and harmful smoke effects than the residents.

\begin{figure}[ht]
    \centering
    \includegraphics[width=0.6\textwidth]{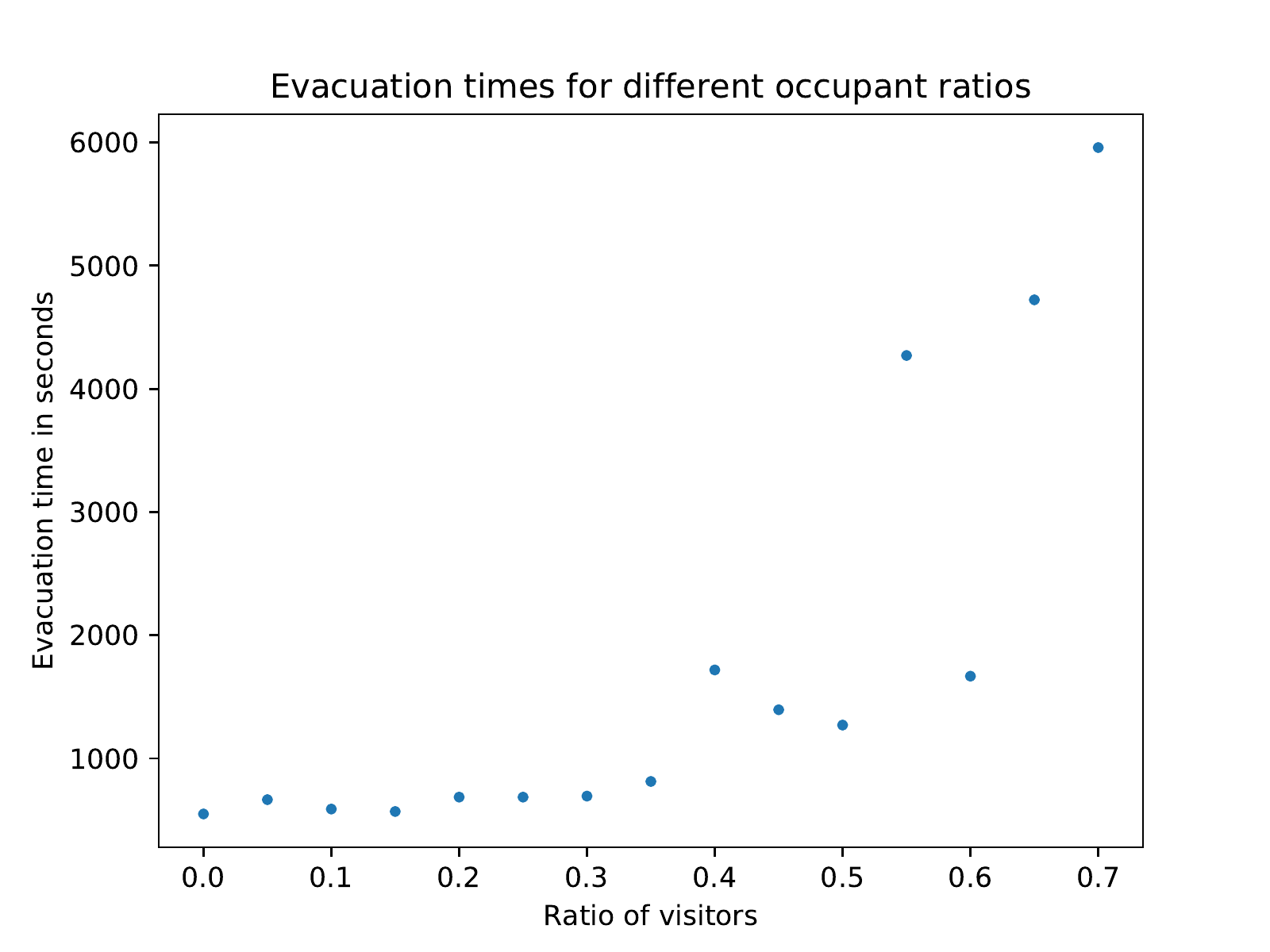}
    \caption{Evacuation time for various values of occupant ratios.}
    \label{fig:evac_to_alpha}
\end{figure}

Finally, we look at the times on which $90\%$ of all evacuees have left the geometry. These times are displayed for several visitor ratios in Figure~\ref{fig:evac_to_alpha}.
We observe a positive trend of increasing evacuation time for increasing values of $\alpha$, but it is apparent that for ratios up to 0.35 the evacuation times are comparable.

It should be pointed out that in this stage of development we do not aim to predict accurate evacuation times; the evacuation times reported here lose credibility for visitor-dominated simulations. This is mainly due to the assumption that visitors without information walk in random directions, without any aid of memory or evacuation signals usually present in buildings. Rather, these simulations try to provide insight in evacuation behaviors for different crowd compositions. 

\section{Conclusions}

The simulations shown in Section \ref{sec:results} draw attention to three fundamental aspects: 
\begin{itemize}
\item A key observation is that evacuations can progress efficiently even when relatively many people have no knowledge of their environment. However, when the number of people that have environment knowledge drops below a certain threshold, evacuation time is strongly impacted. This is in line with conclusions drawn from the experiments in \cite{horiuchi86}, stating that visitors have a significantly longer evacuation time than evacuees familiar with the building.

\item The simulations indicate that having knowledge of the environment increases in some cases the risk to exposure to smoke and fire, possibly due to the fact that no alternatives other than the fastest route to the exit are considered. This is in line with observations made in \cite{bryan99}, which states that people tend to follow their normal patters, even if that comes with additional risks.
Further experimental research is required to draw quantitative conclusions based on these observations.

\item If visibility is reduced, the crowd dynamics becomes driven by social group behaviors. We expect this to involve a high coordination cost and significant information overload (compare the study \cite{Petre}, e.g.).
\end{itemize}

\section{Outlook}
Looking forward, it is our aim to extend the applicability of this simulation framework to cope with more realistic scenarios. Concretely, we would like to improve the smoke simulation by assuming the propagation of the smoke plume follows a coupled thermo-diffusion system interlinked with compressible Navier-Stokes equations, possibly in three dimensions. Additionally, it would be interesting to include the effect of smoke inhalation on evacuees. 

By including more intelligence for visitors, we aim to obtain more realistic evacuation times. By validating these with data on evacuation experiments, we will be able formulate more quantitative conclusions

Finally, for safety reasons, complex buildings or festivals gathering large crowds are endowed with way-signaling systems showing route choices with indications telling people where to go, enhancing their knowledge of a path towards the exit. This brings in the discussion the complexity of social choices.  We have already expanded our simulation framework to cope with way-signaling systems and plan to investigate in the near future their efficiency especially when handling critical pedestrian flow conditions.

\section*{Acknowledgements}
The authors would like to thank Enrico Ronchi (Lund) for useful discussions and the anonymous reviewers for their valuable comments.

\bibliographystyle{ieeetr}      
\bibliography{literature}   

\begin{thebibliography}{10}

\bibitem{vanhees13}
P.~Van~Hees, ``Validation and verification of fire models for fire safety
  engineering,'' {\em Procedia Engineering}, vol.~62, pp.~154--168, 2013.

\bibitem{horiuchi86}
S.~Horiuchi, Y.~Murozaki, and A.~Hukugo, ``A case study of fire and evacuation
  in a multi-purpose office building, {O}saka, {J}apan,'' {\em Fire Safety
  Science}, vol.~1, pp.~523--532, 1986.

\bibitem{pinter17}
N.~Pinter-Wollman, S.~M. Fiore, and G.~Theraulaz, ``The impact of architecture
  on collective behaviour,'' {\em Nature Ecology \& Evolution}, vol.~1, no.~5,
  pp.~s41559--017, 2017.

\bibitem{bae15}
S.~Bae and H.~S. Ryou, ``Development of a smoke effect model for representing
  the psychological pressure from the smoke,'' {\em Safety Science}, vol.~77,
  pp.~57--65, 2015.

\bibitem{cirillo13}
E.~N. Cirillo and A.~Muntean, ``Dynamics of pedestrians in regions with no
  visibility; a lattice model without exclusion,'' {\em Physica A: Statistical
  Mechanics and its Applications}, vol.~392, no.~17, pp.~3578--3588, 2013.

\bibitem{cirillo16}
E.~N. Cirillo, M.~Colangeli, and A.~Muntean, ``Does communication enhance
  pedestrians transport in the dark?,'' {\em Comptes Rendus M{\'e}canique},
  vol.~344, no.~1, pp.~19--23, 2016.

\bibitem{lovreglio16}
R.~Lovreglio, E.~Ronchi, G.~Maragkos, T.~Beji, and B.~Merci, ``A dynamic
  approach for the impact of a toxic gas dispersion hazard considering human
  behaviour and dispersion modelling,'' {\em Journal of {H}azardous
  {M}aterials}, vol.~318, pp.~758--771, 2016.

\bibitem{aube04}
F.~Aub{\'e} and R.~Shield, ``Modeling the effect of leadership on crowd flow
  dynamics,'' {\em Lecture {N}otes in {C}omputer {S}cience}, vol.~3305,
  pp.~601--611, 2004.

\bibitem{pelechano06}
N.~Pelechano and N.~I. Badler, ``Modeling crowd and trained leader behavior
  during building evacuation,'' {\em IEEE {C}omputer {G}raphics and
  {A}pplications}, vol.~26, no.~6, 2006.

\bibitem{tabak10}
V.~Tabak, B.~de~Vries, and J.~Dijkstra, ``Simulation and validation of human
  movement in building spaces,'' {\em Environment and Planning B: Planning and
  Design}, vol.~37, no.~4, pp.~592--609, 2010.

\bibitem{kobes10}
M.~Kobes, I.~Helsloot, B.~De~Vries, and J.~G. Post, ``Building safety and human
  behaviour in fire: A literature review,'' {\em Fire Safety Journal}, vol.~45,
  no.~1, pp.~1--11, 2010.

\bibitem{kobes101}
M.~Kobes, I.~Helsloot, B.~De~Vries, J.~G. Post, N.~Oberij{\'e}, and
  K.~Groenewegen, ``Way finding during fire evacuation; an analysis of
  unannounced fire drills in a hotel at night,'' {\em Building and
  Environment}, vol.~45, no.~3, pp.~537--548, 2010.

\bibitem{jonsson17}
A.~Jonsson, C.~Bonander, F.~Nilson, and F.~Huss, ``The state of the residential
  fire fatality problem in {S}weden: Epidemiology, risk factors, and event
  typologies,'' {\em Journal of Safety Research}, vol.~62, pp.~89--100, 2017.

\bibitem{sorqvist16}
P.~S{\"o}rqvist, ``Grand challenges in environmental psychology,'' {\em
  Frontiers in {P}sychology}, vol.~7, 2016.

\bibitem{becker12}
P.~Becker and M.~Abrahamsson, ``Designing capacity development for disaster
  risk management: A logical framework approach,'' {\em Technical Report},
  2012.

\bibitem{degond13}
P.~Degond, C.~Appert-Rolland, M.~Moussaid, J.~Pettr{\'e}, and G.~Theraulaz, ``A
  hierarchy of heuristic-based models of crowd dynamics,'' {\em Journal of
  Statistical Physics}, vol.~152, no.~6, pp.~1033--1068, 2013.

\bibitem{duong17}
H.~Duong, A.~Muntean, and O.~Richardson, ``Discrete and continuum links to a
  nonlinear coupled transport problem of interacting populations,'' {\em The
  European Physical Journal Special Topics}, pp.~1--13, 2017.

\bibitem{richardson162}
O.~Richardson, ``Large-scale multiscale particle models in inhomogeneuous
  domains: Modelling and implementation,'' Master's thesis, Technische
  Universiteit Eindhoven, 2016.

\bibitem{narain09}
R.~Narain, A.~Golas, S.~Curtis, and M.~C. Lin, ``Aggregate dynamics for dense
  crowd simulation,'' in {\em ACM Transactions on Graphics (TOG)}, vol.~28,
  p.~122, ACM, 2009.

\bibitem{tan15}
L.~Tan, M.~Hu, and H.~Lin, ``Agent-based simulation of building evacuation:
  Combining human behavior with predictable spatial accessibility in a fire
  emergency,'' {\em Information Sciences}, vol.~295, pp.~53--66, 2015.

\bibitem{bryan99}
J.~L. Bryan, ``Human behaviour in fire: the development and maturity of a
  scholarly study area,'' {\em Fire and Materials}, vol.~23, no.~6,
  pp.~249--253, 1999.

\bibitem{mercurial}
O.~Richardson, {\em Mercurial}.
\newblock https://github.com/0mar/mercurial, 2015.
\newblock Python framework for building, running and post-processing crowd
  simulations.

\bibitem{borve15}
S.~Borve, ``{Crowds2D} - a new, robust crowd dynamics simulation model,'' Tech.
  Rep. FFI-rapport 2015/01750, Forsvarets Forskningsinstitutt, 2015.

\bibitem{qi16}
L.~Qi and X.~Hu, ``Design of evacuation strategies with crowd density
  feedback,'' {\em Science China Information Sciences}, vol.~59, no.~1,
  pp.~1--11, 2016.

\bibitem{drysdale2011}
D.~Drysdale, {\em An introduction to fire dynamics}.
\newblock John Wiley \& Sons, 2011.

\bibitem{bystrom2017}
A.~Bystr{\"o}m, {\em Compartment Fire Temperature Calculations and
  Measurements}.
\newblock PhD thesis, Lule{\aa} University of Technology, Structural and Fire
  Engineering, 2017.

\bibitem{smith1996}
E.~E. Smith, ``Heat release rate calorimetry,'' {\em Fire Technology}, vol.~32,
  no.~4, pp.~333--347, 1996.

\bibitem{hirschler91}
M.~M. Hirschler, ``The measurement of smoke in rate of heat release equipment
  in a manner related to fire hazard,'' {\em Fire Safety Journal}, vol.~17,
  no.~3, pp.~239 -- 258, 1991.

\bibitem{cengel97}
Y.~A. Cengel {\em et~al.}, {\em Introduction to {T}hermodynamics and {H}eat
  {T}ransfer}.
\newblock McGraw-Hill New York, 1997.

\bibitem{tewarson08}
A.~Tewarson, ``Smoke emissions in fires,'' {\em Fire Safety Science}, vol.~9,
  pp.~1153--1164, 2008.

\bibitem{mulholland00}
G.~W. Mulholland and C.~Croarkin, ``Specific extinction coefficient of flame
  generated smoke,'' {\em Fire and Materials}, vol.~24, no.~5, pp.~227--230,
  2000.

\bibitem{jin97}
T.~Jin, ``Studies on human behavior and tenability in fire smoke,'' {\em Fire
  Safety Science}, vol.~5, pp.~3--21, 1997.

\bibitem{hughes02}
R.~Hughes, ``A continuum theory for the flow of pedestrians,'' {\em
  Transportation Research Part B: Methodological}, vol.~36, no.~6,
  pp.~507--535, 2002.

\bibitem{treuille06}
A.~Treuille, S.~Cooper, and Z.~Popovic, ``Continuum crowds,'' {\em {ACM} Trans.
  Graph.}, vol.~25, no.~3, pp.~1160--1168, 2006.

\bibitem{cao2014}
S.~Cao, W.~Song, X.~Liu, and N.~Mu, ``Simulation of pedestrian evacuation in a
  room under fire emergency,'' {\em Procedia engineering}, vol.~71,
  pp.~403--409, 2014.

\bibitem{meunier06}
H.~Meunier, J.-B. Leca, J.-L. Deneubourg, and O.~Petit, ``Group movement
  decisions in capuchin monkeys: the utility of an experimental study and a
  mathematical model to explore the relationship between individual and
  collective behaviours,'' {\em Behaviour}, vol.~143, no.~12, pp.~1511--1527,
  2006.

\bibitem{helbing00}
D.~Helbing, I.~Farkas, and T.~Vicsek, ``Simulating dynamical features of escape
  panic,'' {\em Nature}, vol.~407, no.~6803, p.~487, 2000.

\bibitem{cucker07}
F.~Cucker and S.~Smale, ``Emergent behavior in flocks,'' {\em IEEE Transactions
  on automatic control}, vol.~52, no.~5, pp.~852--862, 2007.

\bibitem{moussaid11}
M.~Moussa{\"\i}d, D.~Helbing, and G.~Theraulaz, ``How simple rules determine
  pedestrian behavior and crowd disasters,'' {\em Proceedings of the National
  Academy of Sciences}, vol.~108, no.~17, pp.~6884--6888, 2011.

\bibitem{tsitsiklis95}
J.~N. Tsitsiklis, ``Efficient algorithms for globally optimal trajectories,''
  {\em Automatic Control, IEEE Transactions on}, vol.~40, no.~9,
  pp.~1528--1538, 1995.

\bibitem{kusters17}
R.~Kusters and C.~Storm, ``Dynamic phase separation of confined driven
  particles,'' {\em EPL}, vol.~118, no.~5, p.~58004, 2017.

\bibitem{Petre}
A.~Ciallella, N.~M.~E. Cirillo, and A.~Muntean, ``Free to move or trapped in
  your group: Mathematical modeling of information overload and coordination in
  crowded populations,'' tech. rep., Karlstad University, Sweden, 2018.

\end{thebibliography}


\end{document}